\documentclass[article]{raa} 
\usepackage{setspace, array}
\usepackage{graphicx,times}
\usepackage{natbib}
\usepackage{longtable}
\usepackage{amssymb,amsmath}
\usepackage{pdflscape}
\bibpunct{(}{)}{;}{a}{}{,}
\usepackage{hyperref}
\hypersetup{colorlinks = true, linkcolor = blue, anchorcolor = red, citecolor = blue, filecolor = red, pagecolor = red, urlcolor = blue}

\begin{document}

   \title{Physical and kinematical characteristics of Wolf-Rayet central stars and their host planetary nebulae}

   \volnopage{Vol.0 (20xx) No.0, 000--000}      
   \setcounter{page}{1}          

   \author{Awad, Z \and Ali, A.
      \inst{}
   }

   \institute{Department of Astronomy, Space Science and Meteorology, Faculty of Science, Cairo University, 12613 Giza, Egypt;\\
    {\it zma@sci.cu.edu.eg, afouad@sci.cu.edu.eg}\\
\vs\no
   {\small Received~~2022 month day; accepted~~2023~~May 24}
}

\abstract{We addressed the physical and kinematical properties of Wolf -- Rayet [WR] central stars (CSs) and their hosting planetary nebulae (PNe). The studied sample comprises all [WR] CSs  that are currently known. The analysis is based on recent observations of the parallax, proper motion, and color index of [WR] CSs from the Gaia space mission's early third release (eDR3) catalog, as well as common nebular characteristics.
The results revealed an evolutionary sequence, in terms of decreasing T$_{\text{eff}}$, from the early hot [WO 1] to the late cold [WC 12] stars. This evolutionary sequence extends beyond [WR] CS temperature and luminosity to additional CS and nebular characteristics.
The statistical analysis showed that the mean final stellar mass and evolutionary age of the [WR] CS sample are 0.595 $\pm$ 0.13\,M$_{\odot}$ and 9449 $\pm$ 2437\,yr, respectively, with a mean nebular dynamical age of 7270 $\pm$ 1380\,yr.
In addition, we recognized that the color of the majority ($\sim$ 85\%) of [WR] CSs tends to be red rather than their genuine blue color. The analysis showed that two-thirds of the apparent red color of most [WR]s is attributed to the interstellar extinction whereas the other one-third is due to the PN self-extinction effect.
}

   \authorrunning{Awad \& Ali}            
   \titlerunning{Characterization of [WR] CS and their hosting PNe}  

   \maketitle

%
\section{Introduction}
\label{intro}
Understanding the late stages of stellar evolution provides us with insight into the evolution of our Galaxy since these phases enrich the interstellar medium (ISM) with a lot of elements created inside these stars via nuclear fusion \citep{wil03}. It is now well-established that the evolution of low- and intermediate- mass stars (0.8 $<$ M (M$_{\odot}$) $<$ 8) passes through a transition stage known as planetary nebula (PN). This stage is halfway between the post asymptotic giant branch (post-AGB), in which stars have hydrogen (H) and helium (He) burning shells, and the white dwarf (WD), in which stars have carbon (C) -oxygen (O) cores \citep{Werner&Herwig06}.
Central stars (CS)s of PNe (PNCSs) are often not easy to detect because they are: (1) optically faint ($\sim$ 60\% have V $>$ 15.5 mag), (2) masked by the nebular emission, and sometimes are (3) shifted from the PN center. Based on the H abundance in the atmosphere of CSs, as derived from their spectral analysis, astronomers showed that CSs can fall into two categories: H-rich and H-deficient \citep{Mendez91-1}.
H-deficient CSs show higher abundances of He and C with a small amount of H (or almost H free) compared to their H-rich peers. Stellar spectra of H-deficient CSs are dominated by broad and intense emission lines typical of Wolf-Rayet (WR) massive stars. In order to distinguish WR CSs from the well-known massive WRs, \citet{vandh81} introduced the [WR] notation to denote WR CSs. [WR]s have two main sub-types: carbon-rich [WC], that is dominated by C{\sc iii} 5696\AA~ lines, and oxygen-rich [WO], that is dominated by highly ionized lines of C and O such as C\,{\sc iv} 5806\AA~ and O\,{\sc vi} 3822\AA~ line (e.g. \citealt{weid11}, and references therein).

Early classification of massive WR was qualitative and hence subjective to observers (e.g. \citealt{beals38, hilt66}). The classical WRs were grouped into nitrogen (N) and C stars (WN and WC; respectively) based on the dominant emission features detected in their spectra. According to the ratio between the spectral lines of WN and WC stars, \citet{beals38} divided them into sub-types designed as WN\,5 to WN\,8 and WC\,6 to WC\,8. This classification was then improved by including the line width of the observed emission lines \citep{hilt66}.

\citet{smith90b} reported a new quantitative scheme to define the sub-types WC 4 to WC 9. In this scheme, the principle identifier for WC\,4 - 6 stars is the line ratio C\,{\sc iii} $\lambda$5696 / O\,{\sc v} $\lambda$5590, while for WC\,7 - 9 stars is the line ratio C\,{\sc iv} $\lambda$5808 / C\,{\sc iii} $\lambda$5696. \citet{smith94} suggested a new method to define the WN\,9 - 11 sub-types, while \citet{smith96} used the H abundance and the He\,{\sc ii}\,$\lambda$5411 / He\,{\sc i}\,$\lambda$5875 line ratio to discriminate the ionization subclass of WN\,2 - 9.

\citet{barlow82} proposed that the WO sequence is an evolutionary stage after the WCs and designated the WO sub-types according to the degree of O ionisation to be WO\,1 to WO\,4; from the greatest to the weakest ionized O lines. \citet{king95} defined quantitative spectral type criteria for the whole WO class, and extended the sub-types to include the subclass WO 5 which has an equivalent width of O\,{\sc vi} $\lambda$3811,43 / O\,{\sc v} $\lambda$5590 line $\sim$ 1 compared to 4 in the WO\,4 subclass. Moreover, the authors used the C\,{\sc iii} $\lambda$5696 emission line to distinguish WOs from WC\,4. The classification of WO and WC stars was then revisited by \citet{Crowther98} using high-quality optical observations of twenty WR stars.

\citet{Acker03} have constructed a grid of line intensities normalized to the C{\sc iv} 5806\AA~line and found that their new classification of the hot ([WO\,1-4]) and cool ([WC\,4-11]) CSs is consistent with that of \citet{Crowther98}. This classification is based on the evolution of both the effective temperature (T$_{\text{eff}}$) and stellar winds, where the [WR] evolutionary sequence, in terms of decreasing T$_{\text{eff}}$, can be summarized as [WO\,1]-[WO\,4] $\rightarrow$ [WC 4]-[WC 11].

The existence of He, C, and O absorption lines in the spectra of some H-deficient stars led to the definition of a new class known as PG\,1159, named after the discovery of its prototype star PG\,1159--035 (e.g. \citealt{Napiwotzki&Schoenberner91}; and references therein).

Furthermore, there is a class of CSs with spectra similar to those of [WR] stars but with weaker intensities, narrower line widths, and not required to be of H-poor atmospheres. This class is the so--called weak emission line stars ({\it WELs}; e.g. \citealt{Tylenda93} and \citealt{depew11}). The high resolution and integral field unit (IFU) spectra of numerous PNe hosting {\it WEL} central stars cast doubt on the classification of {\it WELs} \citep{Weidmann15}. The majority of the characteristic spectral lines of {\it WELs} had been reported to be of nebular origin rather than of CS origin (e.g. \citealt{Basurah16, Ali16, Ali19}). Moreover, the results of \citet{misz11} indicated that the classification of {\it WELs} is likely to be attributable to irradiation of the secondary companion in a close binary system.

\citet{Girard07} studied the chemical compositions and other aspects of PNe surrounding {\it WELs}, [WC] and [WO] CSs. The results revealed no substantial variations in the electron temperatures, elemental abundances and the amount of ionizing photons among these types of PNe and the normal PNe. However, they show some differences in their IR properties in which CSs of {\it WELs} class appeared bluer than the other [WR] CSs. In addition, the data analysis supported the hypothesis which proposes an evolutionary sequence from cool [WC\,11] inside dense, low-excitation nebulae to hot [WO\,1] inside low-density, high-excitation nebulae. Also, the results showed no correlation between the PN properties (e.g. the number of ionizing photons, nebular excitation, and electron temperature and density) for PNe associated with {\it WELs} and the other [WR]PNe. This result indicates that {\it WELs} PNe may account for a separate class of objects with a variety of nebular properties that may arise due to confounding effects.

\citet{pena13} examined the circular  (V$_{\text{cir}}$) and radial peculiar (V$_{\text{pec}}$) velocities in samples of {\it WEL}s and [WR] CSs, where they found that [WR]s are concentrated in the Galactic thin-disk more than {\it WELs} and normal PNe indicating a younger population. Furthermore, the vast majority of [WR] nebulae are of Pimbert's type II (V$_{\text{pec}} \sim$ 60 km/s), with a small percentage of type III (V$_{\text{pec}} >$ 60 km/s). These results show that the [WR] phenomena can occur in objects of any stellar mass and age.

Recently, \citet{muth20} used infrared (IR) photometric data to derive the IR properties of three samples of PNe associated with [WR], {\it WELs} , and  normal CSs. The results revealed a tight correlation of decreasing the dust temperature (T$_d$) with the nebular age: ([WCE]: [WO1] to [WC4]) $\rightarrow$ ([WCg]: [WC5] to [WC8]) $\rightarrow$ ([WCL]: [WC 9] to [WC 11]); where [WCE], [WCg], and [WCL] are the early, intermediate, and late types of [WR], respectively. Additionally, the IR luminosity (L$_{\text{IR}}$) of PN show a strong anti-correlation with the PN age and dust mass (m$_d$), while the dust-to-gas mass ratio (m$_d$/m$_g$) does not change noticeably with the PN evolution. More recent studies reported that early types [WR] ([WCE]: [WO1]-[WC5]) have high T$_{\text{eff}}$, ranging from 80,000 K to 150,000 K, while the late types ([WCL]:[WC6]-[WC11]) have T$_{\text{eff}}$ between 20,000 K and 80,000 K (\citealt{dan21}; and  references therein).

\citet{Peimbert78}, introduced a chemical classification for PNe, based on the elemental abundances of He, O, and N relative to H. This classification was revised  by \citet{Quireza07}, where they combined the chemical properties of the PN with some kinematical parameters of the CS such as the Galactic height (Z), V$_{\text {pec}}$, and the initial mass. The results showed that PNe of types I, IIa, and IIb are of population I (pop I) that occupy the Galactic thin-disk while PNe of type III populate the Galactic thick-disk and are of population II (pop II).

Motivated by the huge improvements in the observations of the stellar photometry, distance and proper motion occur in the early third data release ({\it hereafter} eDR3) of the Gaia mission\footnote{Details on Gaia mission: \url{http://sci.esa.int/gaia/}} \citep{Gaia20}, we performed a computational and statistical analysis on a sample of PNe associated with [WR] to derive the main characteristics of both the CS and their hosting nebulae.

This paper is organized as follows: \S~\ref{sample} describes the strategy of selecting and filtering our data sample. The main properties of [WR] CSs and their hosting nebulae are presented in \S~\ref{res} and \ref{wrhost}, respectively, while the origin of [WR] CSs is discussed in \S~\ref{ori}. Finally, we summarize the present work and drew some key conclusions in \S~\ref{conc}.

\section{The data sample}
\label{sample}
The eDR3, which includes the astrometric and photometric data for $\sim 1.8$ billion sources, showed several improvements over the previous second release (DR2; \citealt{Gaia18}). The astrometric parameters include the Galactic positions ($l, b$), proper motion components ($\mu_{\alpha}, \mu_{\delta}$) and the parallax ($\varpi$) while the photometric data includes stellar magnitudes in the visual (G), blue (G$_{\text{BP}}$) and red (G$_{\text{RP}}$) filters for sources as faint as 21 magnitude. The accuracy of the astrometric parameters in eDR3 has been enhanced by a factor of 2 in the proper motion and of $\sim1.5$ in the parallax whereas astrometric errors were reduced by 30 - 40\% for the parallax and by a factor of 2.5 for the proper motion.

In this work, we extracted a sample of PNe associated with [WR] nuclei from the PN CSs catalog of \citet{Weidmann20} which is electronically available, openly, in the Vizier database\footnote{link to \citet{Weidmann20} catalog: \url{https://cdsarc.cds.unistra.fr/viz-bin/cat/J/A+A/640/A10}}.
After collecting the [WR]s sample from the catalogue, we performed a cross-matching between the coordinates of the selected objects and their corresponding values given in both the HASH catalogue\footnote{HASH: \url{http://planetarynebulae.net/EN/hash.php}} and the SIMBAD database\footnote{SIMBAD: \url{https://simbad.u-strasbg.fr/simbad/simbad/simbad/simbad/simbad/simbad/simba}}. Moreover, we  checked the CS position using Aladin Sky Atlas\footnote{Aladin: \url{https://aladin.u-strasbg.fr/}} because a few number of CSs are shifted from their geometric centres in PNe as a result of the interaction with the ISM. Recently, \citet{chorn20} used an automated method to identify the positions of central stars using the Gaia DR2 database. We compared our sample to their catalog and found that only 15 of our sources (out of 120;  12.5\%) are not included in their list. This finding supports the accuracy of the method we adopted in identifying our Gaia eDR3 sources.

\citet{Weidmann20} catalog contains data for $\sim 620$ CSs, collected from numerous literature surveys, including the CS spectral classification as well as the values of surface gravity (g), effective temperature and luminosity (L) for 56.3\%, 69.8\%, 60.5\% of the total number of sources; respectively. The visual magnitude of $\sim$ 74.4\% of the sources is also provided. In this work, the CS luminosity which is listed in \citet{Weidmann20} was updated using the derived distance from the recent Gaia eDR3 parallax. Angular radii, of the entire PN sample, as well as distances for the objects of unknown, negative, and/or highly uncertain parallaxes ($\sigma_{\varpi} \ge$ 50\%) are obtained from \citet{Frew16} while radial velocities are collected from \citet{Durand98} unless stated otherwise. The expansion velocities were collected from \citet{Acker92} except for PNe of unknown measurement, we used the standard value of 20 km/s \citep{Weinberger89}.

A summary of the selected sample, which includes 120 [WR]s classified into: 79 [WC], 38 [WO] and 3 [WN],  is given in Table \ref{tab:1}. The table lists the PN name, [WR] classification, Gaia eDR3 designation, Galactic ($l$, $b$) and Equatorial ($\alpha$, $\delta$) coordinates in degrees. 
In addition, the proper motion in mas/yr, parallax in mas, distance (D) in pc, PN angular radius ($\theta$) in arcsec, radial velocity (V$_r$) in km/s and nebular expansion velocity (V$_{\text {exp}}$) in km/s and their uncertainties are given in Table \ref{tab:2}. It is worth noting that the unique serial number given for each object, in Tables \ref{tab:1} and \ref{tab:2}, has been kept fixed throughout this study. The serial ranges 1-79, 80-117 and 118-120 refer to [WC], [WO] and [WN] sub-classes, respectively.

To  make the results of the current study more precise and informative, we omitted the three [WN] sources from the analysis and focused on the [WO] and [WC] sources due to their large number and their indicated evolutionary sequence \citep{Acker03}. Therefore, the number of sources in our sample is 117.
We divided the data sample into two [WR] groups: early group [WRE] (includes [WO 1] - [WC5]) and late group [WRL] (includes [WC 6] - [WC 12]), following \citet{dan21}. We omitted objects of uncertain and multiple classifications; e.g. objects of class [WC 5/6]. The number of excluded objects is 12, which represents 10\% of the entire [WR] sample.
\begin{longtable}{llllcccc}
\caption{The basic properties of a sample of the entire [WR] studied sample. The complete table is given in Table \ref{tab:1a} in the Appendix.}\\
\hline
{\bf serial} & {\bf PN Name} &{\bf [WR] type}&{\bf Gaia eDR3 Source\_ID}& \multicolumn{2}{c}{\bf {Galactic Coordinates}} & \multicolumn{2}{c}{\bf {Equatorial Coordinates}}\\
 & & & & {\bf l($^{\circ}$)} &{\bf b($^{\circ}$) } &{\bf R.A($^{\circ}$)}  &{\bf Dec.($^{\circ}$)}\\
\hline
\endfirsthead
\multicolumn{8}{c}%
{\tablename\ \thetable\ -- \textit{Continued from previous page}} \\
\hline
{\bf serial} & {\bf PN Name} &{\bf [WR] type}&{\bf Gaia eDR3 Source\_ID}& {\bf l($^{\circ}$)} &{\bf b($^{\circ}$)} &{\bf R.A($^{\circ}$)}  &{\bf Dec.($^{\circ}$)}\\
\hline
\endhead
\hline
\multicolumn{8}{r}{\textit{Continued on next page}} \\
\endfoot
\hline
\endlastfoot
1  &  PN H 1-62  &  [WC 10-11]  &  4045771305065496832  &  0.0100  &  $-$ 6.8504  &  273.3248  &  $-$ 32.3286  \\ [0.5 ex]
2  &  PN M 2-20  &  [WC 5-6]  &  4056503259966429440  &  0.4132  &  $-$ 1.9893  &  268.6055  &  $-$ 29.6023  \\ [0.5 ex]
3  &  PN H 1-47  &  [WC 11]?  &  4062301564840251520  &  1.2949  &  $-$ 3.0402  &  270.1568  &  $-$ 29.3640  
\label{tab:1}
\end{longtable}
\begin{longtable}{llllllll}
\caption{An example of the proper motion ({\bf $\mu_\alpha$}, {\bf $\mu_\delta$ }), parallax ({\bf $\varpi$}), and distance (D) as well as the nebular angular radius ({\bf $\theta$}), radial velocity ({\bf V$_r$}) and expansion velocity ({\bf V$_{\text{exp}}$}) for a sample of the studied entire sample. Distances marked with the $^{\dag}$ symbol are taken from \citet{Frew16}. Full data are listed in Table \ref{tab:2a} in the Appendix.}\\
\hline

{\bf Serial} & {\bf $\mu_\alpha \pm$e }& {\bf $\mu_\delta \pm$e} &{\bf $\varpi \pm$e} &{\bf D$\pm$e} &{\bf $\theta$}  &{\bf V$_r \pm$e} &{\bf V$_{\text{exp}}$} \\
 & {\bf (mas/yr)}& {\bf (mas/yr)} &{\bf (mas)} &{\bf pc}&{\bf (arcmin)}  &{\bf (km/s)} &{\bf (km/s)} \\
[0.5 ex] 
\hline
\endfirsthead
\multicolumn{8}{c}%
{\tablename\ \thetable\ -- \textit{Continued from previous page}} \\
\hline
{\bf Serial} & {\bf $\mu_\alpha \pm$e }& {\bf $\mu_\delta \pm$e} &{\bf $\varpi \pm$e} &{\bf D$\pm$e} &{\bf $\theta$}  &{\bf V$_r \pm$e} &{\bf V$_{\text{exp}}$} \\
 & {\bf (mas/yr)}& {\bf (mas/yr)} &{\bf (mas)} &{\bf pc}&{\bf (arcsec)}  &{\bf (km/s)} &{\bf (km/s)} \\ [0.5 ex]
\hline
\endhead
\hline
\multicolumn{8}{r}{\textit{Continued on next page}} \\
\endfoot
\hline
\endlastfoot
1 & $-$ 3.681 $\pm$ 0.040 & $-$ 3.597 $\pm$ 0.030 & 0.047 $\pm$ 0.034 &  6970 $\pm$ 2430 & 2.25 & $-$ 86.6 $\pm$ 4.7 & 20 \\  [0.5 ex]
2 & 2.778 $\pm$ 0.268 & $-$ 10.904 $\pm$ 0.172 & $-$ 1.930 $\pm$ 0.240 & $^{\dag}$ 7613 $\pm$ 1670 & 1.88 & 58.7 $\pm$ 15.0 & 20 \\  [0.5 ex]
3 & 1.323 $\pm$ 0.063 & $-$ 7.017 $\pm$ 0.044 & 0.091 $\pm$ 0.050 &  9757 $\pm$ 2139 & 1.25 & 107.5 $\pm$ 3.1 & 20 
\label{tab:2}
\end{longtable}
\section{Characterization of [WR] Central Stars.}
\label{res}
\subsection{Physical parameters}
\label{phys-cs}
\begin{figure*}
\begin{center}
\includegraphics[trim=0.6cm 0.1cm 0.7cm 0.7cm,clip=true,width=16cm]{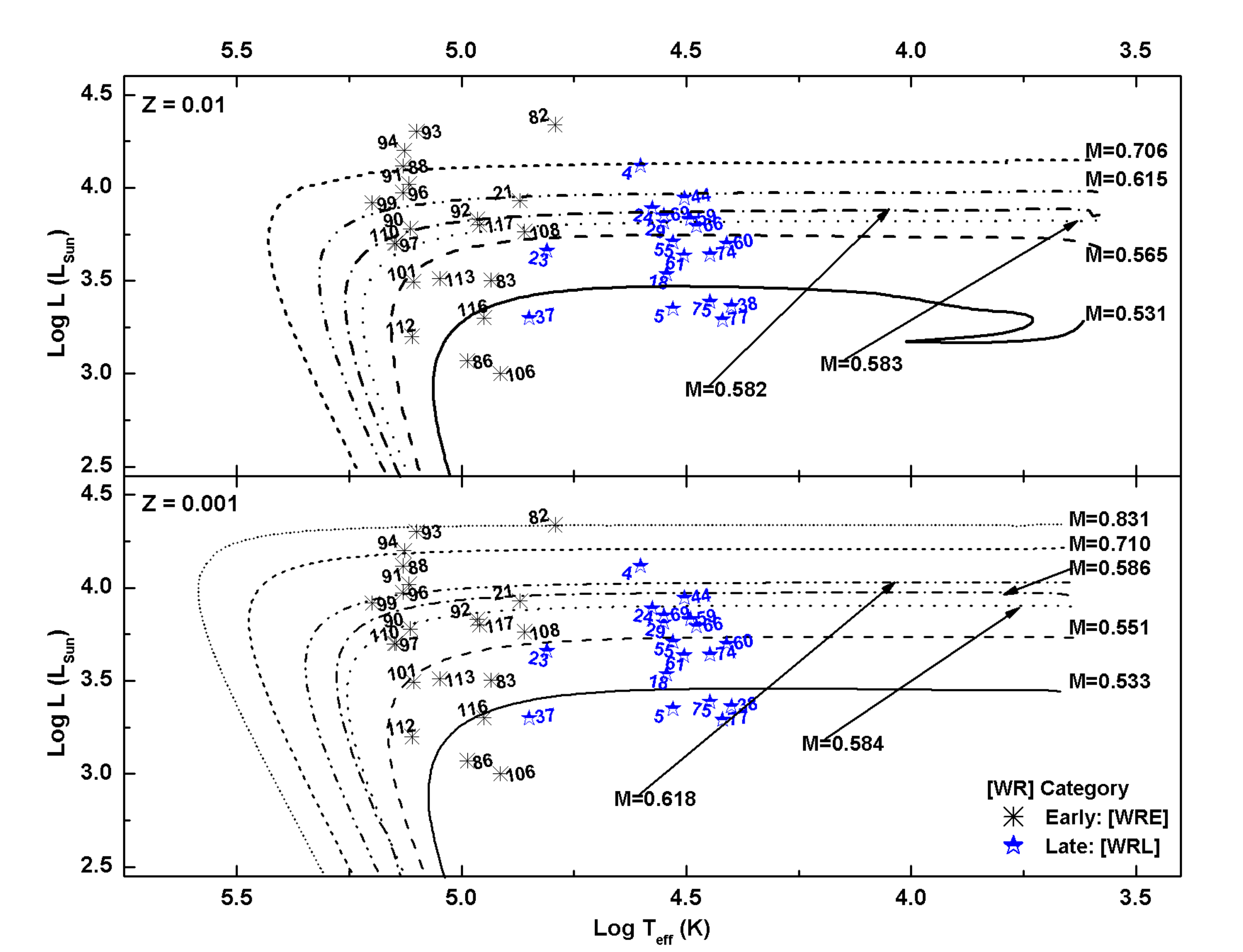}
\caption{The H--R diagram of members of [WRE] (black asterisks) and [WRL] (blue half-filled stars) at two different metalicity values; the solar metalicity (Z = 0.01; top panel) and a lower value of Z = 0.001 (bottom panel). Curves with different line styles represent the final mass isochrones of the CS (see labels) as taken from \citet{Bertolami16}.}
\label{fig:1}
\end{center}
\end{figure*}

Fig. \ref{fig:1} illustrates the Hertzsprung-Russell (H-R) diagram of both [WR] groups at solar metalicity (Z = 0.01; top panel) and a lower metalicity (Z = 0.001; bottom panel), in which the objects are indicated by their serial numbers in Table \ref{tab:1}. [WRE] objects are denoted by black asterisks while [WRL] objects are indicated by blue half-filled stars. In each panel, the CS final mass isochrones, taken from \citet{Bertolami16} models, were indicated using different line styles and labels.

The evolutionary age (t$_{\text{ev}}$ in yr) isochrones were omitted for the figure clarity. It is clear from Fig. \ref{fig:1} that [WR] groups have different locations. [WRE] tends to be concentrated in a zone of higher T$_{\text{eff}}$ with a spread spectrum in their luminosity (L in L$_{\odot}$) compared to [WRL]. 
From the plots, it is noticeable that both groups have a similar general trend of CS final mass distribution, despite the matalicity value. Therefore, the discussion of one of the panels is applicable to the other. In this section, we are discussing the case of low-metalicity (Z = 0.001); displayed in the bottom panel of Fig. \ref{fig:1}.

 For each object in Fig. \ref{fig:1}, we estimated the final mass (M$_{\text{final}}$ in M$_{\odot}$) and the evolutionary age (t$_{\text{ev}}$ in yrs), and listed them with both T$_{\text{eff}}$ (in K) and L (in L$_{\odot}$) in Table \ref{tab:3}. The average value of each computed parameter in the table is recorded at the last row of each group; beneath their corresponding column.

The average T$_{\text{eff}}$ (110.7$\pm$5.9 kK) and L (7554$\pm$1294 L$_{\odot}$) of [WRE]s are higher than those of [WRL]s; 35.8$\pm$2.9 kK and 5223$\pm$661 L$_{\odot}$, respectively. The location of [WRE] and [WRL] group members on the H-R diagram reveals a mean M$_{\text{final}}$ and t$_{\text{ev}}$ of 0.61$\pm$0.08 M$_{\odot}$ and 11012$\pm$3844 yr for [WRE] and $0.58\pm0.07$ M$_{\odot}$ and $7966\pm2749$ yr for [WRL] CSs, respectively. The overall mean M$_{\text{final}}$ and t$_{\text{ev}}$, for the entire sample, are $0.595\pm0.013$ M$_{\odot}$ and $7966\pm2749$ yr, respectively. The mean [WR] M$_{\text{final}}$ is higher than that of hydrogen-poor PG1159 CS class ($0.58\pm0.07$ M$_{\odot}$) while the mean [WR] t$_{\text{ev}}$ is smaller than that of PG1159 CS class ($25500\pm5300$ yr; \citealt{Ali21}). According to \citet{Bertolami16}, the average M$_{\text{final}}$ of [WRE] and [WRL] corresponds to an initial stellar mass of 2.5 M$_{\odot}$ and 1.5 M$_{\odot}$, respectively. Moreover, the mean t$_{\text{ev}}$ of [WRE] and [WRL] is $11012\pm3844$ yr and $7966\pm2749$ yr, respectively.

\subsection{Kinematical parameters}
\label{kin-char}
The kinematical characteristics of [WR]s are analysed by investigating the Galactic velocity distribution of the stars through constructing the “Toomre diagram” and by computing their radial peculiar velocity. In this study we utilized both techniques.
We calculated the space velocity components (U, V, W) and their uncertainties for early and late [WR] members following \citet{Johnson&Soderblom87}; provided that the [WR] proper motion, parallax and radial velocity are known. The `U' and `V' components denote the Galactic center and Galactic rotation directions, respectively, whereas the component `W' indicates the direction perpendicular to the Galactic disk. These components are corrected to the local standard of rest (LSR). This correction yields the (U$_{\text {LSR}}$, V$_{\text {LSR}}$, W$_{\text {LSR}}$) components by adding the Solar motion components (U$_{\odot}$, V$_{\odot}$, W$_{\odot}$) adopted to be (10, 5.5, 7.17) following \citet{Dehnen&Binney98}. The corrected space rotational velocity component `V$_{\text {LSR}}$' has been further corrected to the solar motion around the galactic center by adopting a solar rotational velocity (V$_{R}$ = 220 km/s). Using (U$_{\text {LSR}}$, V$_{\text {LSR}}$, W$_{\text {LSR}}$), we computed the total space velocity, V$_s$, (= $\sqrt{U^2_{\text {LSR}} + V^2_{\text {LSR}} + W^2_{\text {LSR}}}$).  The V$_{\text{pec}}$ is calculated followed the procedure given by \citet{Quireza07}, assuming that the Galactocentric distance of the Sun is $R_{\odot} = 7.6$ kpc and the solar rotational velocity is V$_{R}$ = 220 km/s. The uncertainty in V$_s$ is estimated by propagating the errors in U, V, and W components. The U$_{\text {LSR}}$, V$_{\text {LSR}}$, W$_{\text {LSR}}$, V$_s$, and V$_{\text{pec}}$ (all in km/s), for both [WRE] and [WRL] groups, are listed in Table \ref{tab:3}.
\begin{landscape}
\begingroup
\footnotesize
\begin{longtable}{lllllllllllll}
\caption{The physical and kinematical parameters ofa few number of objects in our [WR] sample organized in two groups [WRE] and [WRL]. Full data is available in the Appendix in Table \ref{tab:3a}.}\\
\hline
{\bf serial} & {\bf PN Name} & {\bf [WR] type}& \multicolumn{4}{c}{\bf Physical properties}&&\multicolumn{5}{c}{\bf Kinematical Parameters}\\

&&&{\bf log T$_{\text{eff}}$} & {\bf log L}& {\bf M$_{\text{final}}$ }& {\bf t$_{\text{ev}}$} & &{\bf U$_{\text{LSR}} \pm$ e} &{\bf V$_{\text{LSR}} \pm$ e}& {\bf W$_{\text{LSR}} \pm$ e}& {\bf V$_s \pm$ e} &{\bf V$_{\text{pec}} \pm$ e}\\
 & &&{\bf (K)}& {\bf (L$_{\odot}$)} &{\bf (M$_{\odot}$)} &{\bf (yr)}&& \multicolumn{5}{c}{\bf km/s}\\ [0.5 ex] 
\hline
\endfirsthead
\multicolumn{12}{c}%
{\tablename\ \thetable\ -- \textit{Continued from previous page}} \\
\hline
{\bf serial} & {\bf PN Name} & {\bf [WR] type}&{\bf log T$_{\text{eff}}$} & {\bf log L }& {\bf M$_{\text{final}}$ }& {\bf t$_{\text{ev}}$} & &{\bf U$_{\text{LSR}} \pm$ e} &{\bf V$_{\text{LSR}} \pm$ e}& {\bf W$_{\text{LSR}} \pm$ e}& {\bf V$_s \pm$ e} &{\bf V$_{\text{pec}} \pm$ e} \\
\hline
\endhead
\hline
\multicolumn{12}{r}{\textit{Continued on next page}} \\
\endfoot
\hline
\endlastfoot
\multicolumn{12}{c}{\bf [WRE] }  \\ [0.5 ex] \hline

 7 & PN K 5-3 & [WC 4] &  &  &  &  & & 168.7 $\pm$ 8.3 & 8.4 $\pm$ 2.0 & $-$ 13.8 $\pm$ 8.5 & 271 $\pm$ 5.4 & 127 $\pm$ 7.6 \\  [0.5 ex]
 11 & PN M 2-31 & [WC 4] &  &  &  &  & &    &    &    &    & 89 $\pm$ 14.1 \\  [0.5 ex]
 12 & PN M 3-15 & [WC 4] &  &  &  &  & & 117.5 $\pm$ 2.7 & 185.4 $\pm$ 52.5 & $-$ 61.7 $\pm$ 14.4 & 137 $\pm$ 7.3 & 47 $\pm$ 11.5 \\  [0.5 ex]
 $\vdots$ & &  &  &  & $\vdots$ & $\vdots$& &    &    &    & $\vdots$ & $\vdots$ \\  [0.5 ex]
\hline
  &{\bf Average} &  &  &  & 0.61 & 11012 & &    &    &    & 115 $\pm$ 25.8 & 51 $\pm$ 11.8 \\  [0.5 ex] \hline
\multicolumn{12}{c}{\bf [WRL] }  \\ [0.5 ex] \hline
 1 & PN H 1-62 & [WC 10-11] &  &  &  &  & & $-$ 68.1 $\pm$ 4.7 & 63.2 $\pm$ 22.0 & 70.1 $\pm$ 20.1 & 185 $\pm$ 20.2 & 73 $\pm$ 3.9 \\  [0.5 ex]
 4 & PN SwSt 1 & [WC 9]pec & 4.60 & 4.12 & 0.65 & 212 & & 3.5 $\pm$ 2.0 & 164.2 $\pm$ 52.3 & 74.1 $\pm$ 23.1 & 93 $\pm$ 36.4 & 10 $\pm$ 1.3 \\  [0.5 ex]
 5 & PN H 1-55 & [WC 11] & 4.53 & 3.35 & 0.53 & 38139 & & $-$ 14.2 $\pm$ 19.3 & 227.5 $\pm$ 207.1 & 138.9 $\pm$ 58.8 & 140 $\pm$ 59.5 & 93 $\pm$ 38.1 \\  [0.5 ex]
 $\vdots$ & &  &  &  & $\vdots$ & $\vdots$& &    &    &    & $\vdots$ & $\vdots$ \\  [0.5 ex]
\hline
  & {\bf Average} &  &  &  & 0.58 & 7966 & &    &    &    & 129 $\pm$ 29.2 & 59 $\pm$ 15.9 
\label{tab:3}
\end{longtable}
\endgroup
\end{landscape}
\subsubsection{Toomre diagram}
\label{tom}
\begin{figure}
\begin{center}
\includegraphics[trim=0.0cm 0.0cm 0.0cm 0.0cm,clip=true,width=12cm]{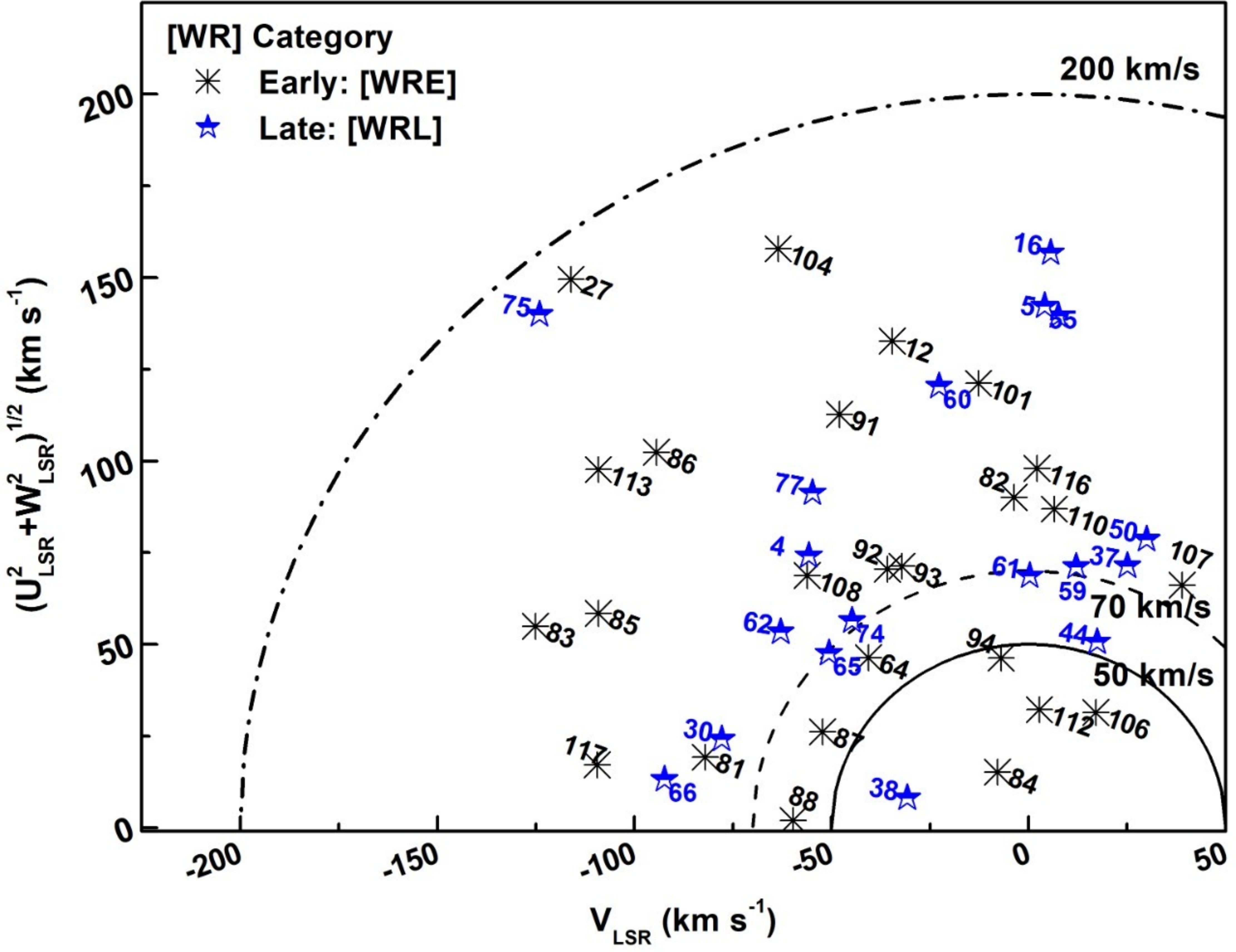}
\caption {The Toomre diagram for [WRE] and [WRL] group members (see figure key). The solid, dashed, and dash-dotted semi-circles refer to the constant total space velocity values at 50, 70 and 200 km/s, respectively. The error bars are omitted for the figure clarity.}
\label{fig:2}
\end{center}
\end{figure}

The Toomre diagram describes the stellar distribution within the Galactic components (the thin-disk, the thick-disk, and the halo) according to their space velocity components \citep{Bensby03, Bensby10}. The graph is a relation between the velocity components $\sqrt{\text{U}^2_{\text{LSR}}~+~\text{W}^2_{\text{LSR}}}$ and V$^2_{\text{LSR}}$ (see Fig. \ref{fig:2}). Stellar locations within the Galactic components are defined by their total V$_s$ as follows:

$$
\mbox{If V$_s$ (km s$^{-1}$)}
\begin{cases}
< ~ 50 \quad \Longrightarrow  \mbox{the Galactic thin-disc} \\
50 ~ \leq~ \mbox{V$_s$}~ \leq ~ 70 \quad \Longrightarrow \mbox{either the Galactic thin- or thick-disc} \\
70 ~ \leq~ \mbox{V$_s$}~ \leq ~ 200 \quad \Longrightarrow \mbox{the Galactic thick-disc} \\
> ~ 200 \quad \Longrightarrow  \mbox{the Galactic halo} \\
\end{cases}
$$

Fig. \ref{fig:2} displays the Toomre diagram of [WRE] and [WRL] group members. Semi-circles indicate constant V$_s$ at 50, 70 and 200 km/s, see labels, taken from \citet{Bensby03, Bensby10}. From the distribution of these objects, we noticed that 74\% of the [WRE] and 80\% of [WRL] members belong to the Galactic thick-disc while only 13\% and 5\% of the [WRE] and [WRL] members, respectively, are linked to the Galactic thin-disk. The rest of the studied sample possesses 50 $<$ V$_s <$ 70 which implies that these objects reside within the overlapping region between the thin- and thick- disks. The results indicate that the majority of [WR] CS are associated with the Galactic thick-disk and there is no discrepancy between the [WRE] and [WRL] groups in terms of their location. 

\subsubsection{Radial peculiar velocity}
\label{pec}
\citet{pena13} analysed the V$_{\text {pec}}$ of [WR], {\it WELs}, and normal CSs, where they found that most of the PNe associated with [WR] are of Peimbert type II with V$_{\text {pec}} \le 60$ km/s and a minor fraction of type III with V$_{\text {pec}} > 60$ km/s. Additionally, they compared the kinematical properties of the [WR] and {\it WELs} samples and concluded that PNe associate with [WR] and those accompanied with {\it WEL} stars are not related.

We computed the V$_{\text {pec}}$ for all members in the two [WR] groups and found that the mean absolute V$_{\text {pec}}$ of [WRE] and [WRL] are 51$\pm$12 and 58$\pm$16, respectively; see Table \ref{tab:3}. The analysis showed that about 70\% and 57\% of [WRE] and [WRL] members, respectively, are located in the Galactic thin-disk, and hence they are of pop I, while the rest of the [WR] CSs occupies the Galactic thick-disk. Our findings are in agreement with the results of \citet{pena13}, but they apparently contradict the results we obtained from the Toomre diagram. 
To test whether or not this contradiction is true, we performed a t-test (a statistical test aims to determine whether two processes or two groups are different) for our data. The results of the t-test revealed that the statistical significance level is very low (0.001), indicating that the two results are very close to each other and that the apparent disagreement is most likely due to the obtained uncertainties ($\sim$ 20\%) in both V$_s$ and V$_{\text{pec}}$.

\section{Characterizing the [WR] hosting nebulae}
\label{wrhost}
\subsection{Physical properties}
\label{phys-pn}

Table \ref{tab:4} summarizes the main physical characteristics of the PNe associated with [WR]s. The results show that nebular shells around [WRE]s possess a mean larger size (R$= 0.164\pm 0.03$ pc) and dynamical age (t$_{\text {dyn}} = 7270\pm1380$ yrs) compared to those associated with [WRL] CSs: R = 0.111$\pm$0.024 pc and t$_{\text {dyn}}$ = 5350$\pm1170$ yr.

The nebular 5 GHz emission is an essential factor for determining some physical properties such as the ionized mass (M$_{\text{ion}}$), optical depth ($\tau$) and radio surface temperature (T$_{\text{B}}$). The mean value of the 5 GHz emissions unveiled that [WRE]s emit more radiation ($163\pm16$ mJy) than [WRL]s ($100\pm10$ mJy). We derived a mean ionized mass of $0.199\pm0.099$ M$_{\odot}$ and $0.076\pm0.029$ M$_{\odot}$ for [WRE] and [WRL], respectively. This result is consistent with the results of the initial/final mass function (e.g. \citealt{Jeffries97}) and theoretical modeling predictions (e.g. \citealt{Marigo01}), where the nebular ionized mass is expected to be larger for nebulae that arise from massive progenitor stars.

The T$_{\text{B}}$ parameter of PN shells can be used as a sign for their different evolutionary stages because it depends on the nebular expansion rate, mass and the exact location of the CS within the H--R plane \citep{Phillips05}. Our results showed that [WRL] has higher T$_{\text{B}}$ (580\,K) than [WRE] CSs (T$_{\text{B}}$ = 106\,K). \citet{Cahn92} suggested that the T$_{\text{B}}$ parameter is linked to the PN optical depth ($\tau$), where large value of $\tau$ corresponds to low optical depth (i.e. optically thin gas) whereas small value of $\tau$ corresponds to optically thick gas. T$_B$ increases from optically thin ($\tau >$ 3.13) to optically thick ($\tau <$ 3.13) nebular shells, where $\tau \sim$ 3.13 indicates the transition phase from optically thick to optically thin.
The mean value of $\tau$ implies that gas shells of PNe associated with [WRL] CSs are optically thicker than those hosting [WRE] CSs.

The results, in Table \ref{tab:4}, suggest an evolutionary sequence from early to late for PNe similar to that of their [WR] stars. In that, early PNe phases are associated with higher 5 GHz emission and larger ionized mass than late PNe phases of PNe. The average values of T$_{\text{B}}$ and $\tau$ reveal a tendency for PN shells around [WRE] to be optically thinner and cooler compared to PN shells around [WRL] stars.
\begingroup
\footnotesize
\begin{longtable}{lllllllll}
\caption{A portion of the physical nebular parameters surrounding the [WR]s. The PN scale height (Z), radius (R), 5GHz flux ({\bf F$_{\text{5GHz}}$}), ionized mass ({\bf m$_{ion}$}), optical depth ({\bf $\tau$}), and radio surface temperature ({\bf T$_B$}). The entire table is available as Table \ref{tab:4a} in the Appendix.}\\
\hline
{\bf serial} & {\bf PN Name} &{\bf Z $\pm$ e }& {\bf R $\pm$ e} &{\bf t$_{\text{dyn}} \pm$ e}&{\bf F$_{\text{5GHz}}\pm$ e} & {\bf m$_{ion}\pm$ e} & {\bf $\tau$} &{\bf T$_B$} \\
             &               &{\bf (pc)}      & {\bf (pc)}      &{\bf (yr/10$^3$)}& {\bf (mJy)}& (M$_{\odot}$) & &{\bf (K)}\\ [0.5 ex]
\hline
\endfirsthead
\multicolumn{9}{c}%
{\tablename\ \thetable\ -- \textit{Continued from previous page}} \\
\hline
{\bf serial} & {\bf PN Name} &{\bf Z $\pm$ e }& {\bf R $\pm$ e} &{\bf t$_{\text{dyn}} \pm$ e}&{\bf 5GHz $\pm$ e}& {\bf m$_{ion} \pm$ e} & {\bf $\tau$} &{\bf T$_B$} \\
             &               &{\bf (pc)}      & {\bf (pc)}      &{\bf (yr/10$^3$)}& {\bf (mJy)}& & &{\bf (K)} \\ [0.5 ex]
\hline
\endhead
\hline
\multicolumn{7}{r}{\textit{Continued on next page}} \\
\endfoot
\endlastfoot
\multicolumn{9}{c}{\bf [WRE] }  \\ [0.5 ex] \hline
 7 & PN K 5-3 & 443 $\pm$ 97 & 0.143 $\pm$ 0.032 & 7.00E+03 $\pm$ 1.54E+03 &    &    &  &  \\ [0.5 ex]
 11 & PN M 2-31 & 383 $\pm$ 75 & 0.057 $\pm$ 0.011 & 2.77E+03 $\pm$ 5.44E+02 & 51.0 $\pm$ 10.20 & 0.069 $\pm$ 0.036 & 2.46 & 252 \\ [0.5 ex]
 12 & PN M 3-15 & 391 $\pm$ 81 & 0.057 $\pm$ 0.012 & 2.78E+03 $\pm$ 5.75E+02 &    &    &  &  \\ [0.5 ex]
$\vdots$ & & $\vdots$ & $\vdots$ & $\vdots$ & $\vdots$ & $\vdots$ & $\vdots$ & $\vdots$ \\ [0.5 ex]
\hline
  & {\bf Average} & 370 $\pm$ 75 & 0.164 $\pm$ 0.030 & 7.27E+03 $\pm$ 1.38E+03 & 162.5 $\pm$ 16.15 & 0.199 $\pm$ 0.099 & 3.33 & 106 \\ [0.5 ex]\hline
 \multicolumn{9}{c}{\bf [WRL] }  \\ [0.5 ex] \hline
 1 & PN H 1-62 & 831 $\pm$ 290 & 0.076 $\pm$ 0.027 & 3.72E+03 $\pm$ 1.30E+03 &    &    &  &  \\ [0.5 ex]
 4 & PN SwSt 1 & 341 $\pm$ 109 & 0.038 $\pm$ 0.012 & 2.71E+03 $\pm$ 8.70E+02 & 184.3 $\pm$ 6.00 & 0.035 $\pm$ 0.028 & 2.20 & 463 \\ [0.5 ex]
 5 & PN H 1-55 & 749 $\pm$ 259 & 0.068 $\pm$ 0.023 & 3.31E+03 $\pm$ 1.15E+03 &    &    &  &  \\ [0.5 ex]
$\vdots$ & & $\vdots$ & $\vdots$ & $\vdots$ & $\vdots$ & $\vdots$ & $\vdots$ & $\vdots$ \\ [0.5 ex]\hline
  & {\bf Average} & 418 $\pm$ 96 & 0.111 $\pm$ 0.024 & 5.35E+03 $\pm$ 1.17E+03 & 99.7 $\pm$ 9.97 & 0.076 $\pm$ 0.029 & 2.57 & 580 \\ [0.5 ex]\hline
\label{tab:4}
\end{longtable}
\endgroup
\subsection{PN morphology, excitation class, and classification}
\label{mor}
While higher masses of PN progenitor stars are expected to give eject both bipolar (B) and elliptical (E) outflows, circular (round; R) PNe are most likely to emanate from Sun-like stars with masses 1 - 1.2 M$_{\odot}$. The bipolar sources appear to come from far more massive stars, may be belonging to spectral type A3 or earlier \citep{Phillips04}. There are no differences obtained for the distribution of the three well-known morphological PN types among early and late [WR] nebulae. Both BPNe and EPNe types account for around 40\% of the entire studied [WR] CS sample, whereas RPNe account for the rest of the objects ($\sim$ 20\%). Table \ref{tab:5} summarizes the basic morphological classes, excitation class (EC), and Peimbert classification of the [WR] nebulae. 

The excitation class of PNe is an important parameter that associates with few PN factors such as the nebular structure, mass, chemical abundances, and the temperature and luminosity of the CS \citep{Gurzadian91}. The results reveal that [WRE] nebulae have a higher mean EC (6.9) than [WRL] nebulae (1.9). This result is consistent with the higher ionized mass, effective temperature and luminosity of the early [WR] group compared to the late [WR] group. Further, we compared the mean EC of [WRE] PNe with the non-[WR], {\it WEL}, and [WR] CSs that given by \citet{muth20}, where the result shows a slightly higher value than non-[WR](EC = 6.6), and greater than {\it WEL} (EC = 5.2) and [WR] (EC = 4.3) nebulae.

The results also reveal that 60\% of [WRE] nebulae are of Peimbert type IIa, whereas 46\% of [WRL] nebulae are of Peimbert type IIb. Moreover, Type III members account only for 12\% of [WRE] and 36\% of [WRL] PNe. This finding is in agreement with the idea that [WRE] nebulae originate from more massive progenitor stars than [WRL] nebulae. This result in addition to the mean scale height (Z) and V$_{\text{pec}}$ of both [WR] groups indicate that [WRE] PNe have the tendency to belong to the Galactic thin-disk while [WRL] nebulae tend to reside in the Galactic thick-disk.
\subsection{Infrared properties of [WR] nebulae}
\label{IR}

While PNe shells are mostly ionized, there is an evidence that they contain significant amounts of molecular material and dust grains. Visual extinction (A$_v$), especially in bipolar PNe, and the excess emission in infrared (IR) wavelengths indicate the presence of dust grains.
\citet{Gorny01} examined the IR characteristics of 49 [WR] PNe, where they found that the mean T$_d$, increases from early to late [WR] types. Furthermore, \citet{Gorny01} suggested that [WR] PNe form a homogeneous class of an evolutionary sequence, similar to their associated CSs, from compact nebulae around [WRL] stars to the more diffuse nebulae surrounding [WRE] stars.

\citet{muth20} investigated the IR properties of $\sim$ 100 [WR] PNe, and they found the IR characteristics of [WR] PNe are comparable to those surrounding normal and {\it WEL} CSs. The results showed that there is a tight anti-correlation between the age of the PN, and both the T$_d$ and the infrared luminosity (L$_{\text{IR}}$). The study, also, revealed that L$_{\text{IR}}$ of PNe has another strong anti-correlation with PN dust mass (m$_d$) and dust-to-gas mass ratio  (m$_d$/m$_g$), in which both parameters do not change noticeably with nebular evolution time. Moreover, \citet{muth20} found that the m$_d$/m$_g$ ratio is not correlated with the age of the nebula. This result supports the previous findings of \citet{Gorny01}. 

\begingroup
\footnotesize
\begin{longtable}{lllllllll}
\caption{The excitation class (EC), Peimbert (Peim.) and morphological (mor.) types, dust temperature ({\bf T$_{\text{dust}}$} ), dust mass ({\bf m$_{\text{dust}}$}), dust/gas mass ratio ({\bf m$_{\text{dust}}$/m$_{\text{gas}}$ }), and infrared luminosity ({\bf L$_{\text{IF}}$}) of [WR]PNe. The entire table is available in the Appendix in Table \ref{tab:5a}. }\\
\hline
{\bf serial} & {\bf PN Name} & {\bf EC } & {\bf Peim.} & {\bf Mor.}  & {\bf T$_{\text{dust}}$} & {\bf m$_{\text{dust}} \pm$ e}& {\bf m$_{\text{dust}}$/m$_{\text{gas}} \pm$ e}& {\bf L$_{\text{IF}} \pm$ e} \\
& &  & {\bf Type} & {\bf Type}  & (K) & ($10^{-4}$\,M$_{\odot}$)&($10^{-3}$) &  (L$_{\odot}$) \\ [0.5ex]
\hline
\endfirsthead
\multicolumn{9}{c}%
{\tablename\ \thetable\ -- \textit{Continued from previous page}} \\
\hline
{\bf serial} & {\bf PN Name} & {\bf EC } & {\bf Peim.} & {\bf Mor. }  & {\bf T$_{\text{dust}}$} & {\bf m$_{\text{dust}} \pm$ e}& {\bf m$_{\text{dust}}$/m$_{\text{gas}} \pm$ e}& {\bf L$_{\text{IF}} \pm$ e} \\
& &  & {\bf Type} & {\bf type} & (K) & (M$_{\odot}$)& &  (L$_{\odot}$) \\ [0.5ex]
\hline
\endhead
\hline
\multicolumn{9}{r}{\textit{Continued on next page}} \\
\endfoot
\endlastfoot
\multicolumn{9}{c}{\bf [WRE] }  \\ [0.5 ex] \hline
 8 & Hen 2-436 & 3.0 &  & S &  115 & 1.84 $\pm$ 0.75 &    & 2934 $\pm$ 1250 \\  [0.5 ex]
 11 & PN M 2-31 & 2.0 &  & E  & 86 & 1.73 $\pm$ 0.72 & 2.9 $\pm$ 0.5 & 642 $\pm$ 340 \\  [0.5 ex]
  $\vdots$ & & $\vdots$  &  & &  $\vdots$  & $\vdots$ & $\vdots$  & $\vdots$ \\ [0.5 ex]
\hline
  & {\bf Average} & 6.8 &  &  & 86.4 & 1.2 $\pm$ 0.5 & 3.9 $\pm$ 0.7 & 640.0 $\pm$ 283.6 \\ [0.5 ex]\hline
\multicolumn{9}{c}{\bf [WRL] }  \\ [0.5 ex] \hline
 4 & PN SwSt 1 & 8.0 & IIb & S &  170 & 0.52 $\pm$ 0.32 &    & 5825 $\pm$ 2830 \\  [0.5 ex]
 5 & PN H 1-55 & 1.0 & IIb & B &  73 & 5.42 $\pm$ 2.47 &    & 890 $\pm$ 436 \\  [0.5 ex]
 $\vdots$ & & $\vdots$  &  &  &  $\vdots$  & $\vdots$ & $\vdots$  & $\vdots$ \\ [0.5 ex]\hline
  & {\bf Average} & 1.9 &  &  &  114.1 & 5.2 $\pm$ 2.4 & 12.0 $\pm$ 2.1 & 4310.6 $\pm$ 2129.6 \\  [0.5 ex]\hline
\multicolumn{6}{l}{Morphological types are: R:Round, E: Elliptical, B: Bipolar, S: Stellar }\\
\label{tab:5}
\end{longtable}
\endgroup
The results presented in Table \ref{tab:5} show an increase in the mean T$_d$ (similar to T$_{\text{B}}$) from [WRE] to [WRL] nebulae. This finding is consistent with the results of \citet{Gorny01}. In addition, the mean value of m$_d$, m$_d$/m$_g$, and L$_{IR}$ parameters decreases from [WRL] to [WRE] PNe. This may indicate that a large fraction of dust has been destroyed by the high amount of UV radiation that emitted from the CS during the early phase of [WR] stars. All the IR mean parameters, listed in Table \ref{tab:5}, decrease with increasing the nebular age providing further evidence for the [WR] evolutionary sequence from late to early types. The mean value for the three IR parameters for [WRL] PNe are higher than those of {\it WEL}s and normal PN CSs.
\section{Origin of [WR] central stars}
\label{ori}
The origin of both [WR] and {\it WEL}s central stars in PNe is not, yet, well-understood. Different hypotheses were proposed to explain the formation of [WR] suggest that these stars originate either from the direct evolution of AGB (through a born--again scenario) or in a close binary system that goes through a common envelope phase (e.g. \citealt{Gorny01, Crowther06}; and references therein). 
Studies suggest that the occurrence of the thermal pulse (TP) in the post-AGB phase can result in an H-deficient CS (e.g. \citealt{iben83, iben84}; \citealt{herw99}; \citealt{depew-phd} and \citealt{Danehkar14A}). The surface abundances of H-deficient star is strongly dependent on the occurrence time of the TP beyond the AGB phase: (1) The AGB final thermal pulse (AFTP) happens at the end of the AGB when the envelope has extremely low mass and the CS has not yet entered the PN phase. This mechanism reduces the H surface abundance (15\% by mass), but increases the abundance of both C and O. As a result, it cannot account for the normal surface abundances of [WR] stars. (2) The late thermal pulse (LTP) takes place when the star advances with constant luminosity from the AGB to the white dwarf (WD) phase, and the CS has recently evolved through a PN phase. The H surface abundance remains constant during the thermal pulse. Only when the star returns to the AGB phase, the so-called born-again event occurs and a dredge-up mixing process reduces hydrogen at the surface to a few percent ($\leq 5\% $) by mass; (3) When the star is on the WD cooling track, a very late thermal pulse (VLTP) occurs. During this TP, the H surface abundance is entirely blended into the hot inner layers and the convective H--burning generates a hydrogen free star and the star is then returned to the AGB phase. The born-again scenario is suggested to be the most likely explanation for the birth of [WR] CS. Furthermore, the VLTP event has two returns to the AGB; the first returns rapidly in a few years, whilst the second takes a little longer, approximately $10^2$ yr.

\begin{figure}
\begin{center}
\includegraphics[trim=0.4cm 0.2cm 0.5cm 0.5cm,clip=true,width=13cm]{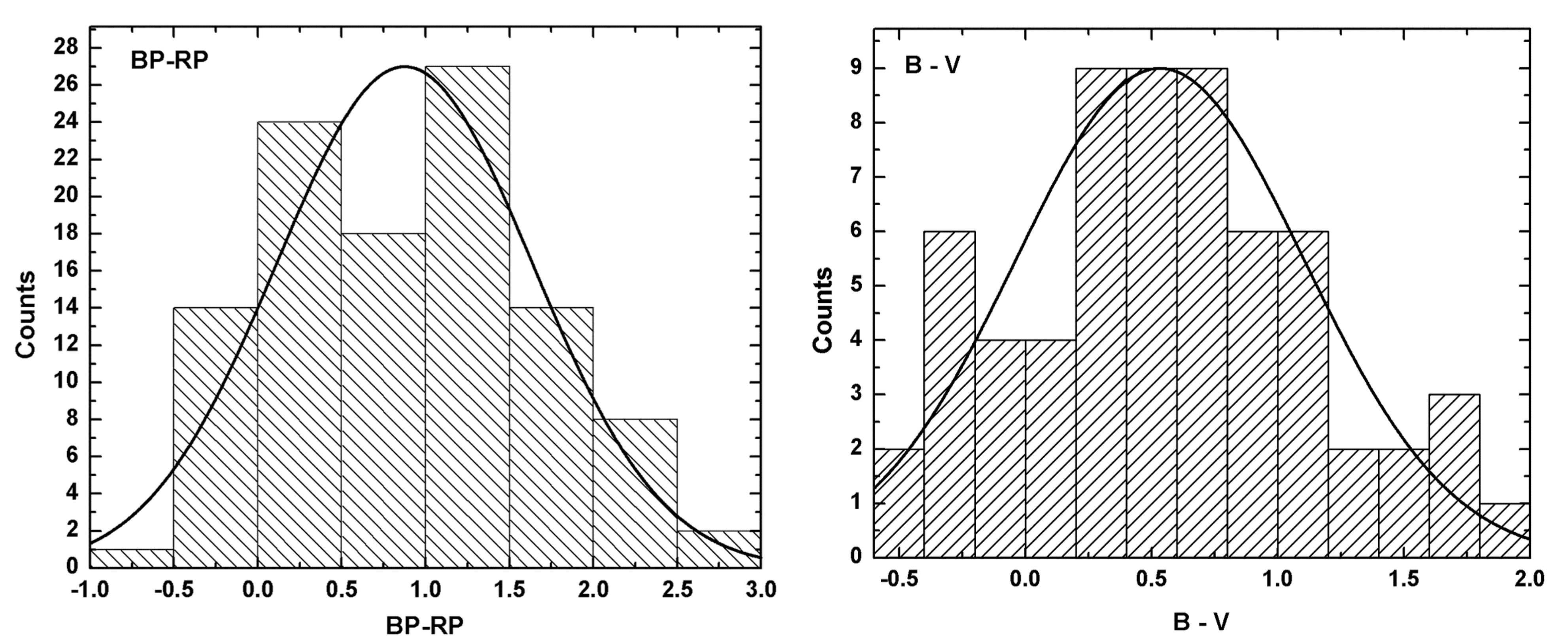}
\caption {A histogram represents the observed color distribution of our sources. left: (BP-BR) from Gaia eDR3 database, right: (B-V) taken from \citet{Acker92}.}
\label{fig:3}
\end{center}
\end{figure}

Unlike normal PN CSs, which are commonly of blue color, the majority of our  sample (85\%) tends to be red; i.e. (BP-RP) $>$ 0.0; see column 4 in Table \ref{tab:6}. To confirm this result, we collected the B and V magnitudes from \citet{Acker92}, where we attained the same result. Figure \ref{fig:3} shows the frequency distributions of (BP-RP) and (B-V) indices, which show that the mean values of (BP-RP) and (B-V) are roughly 0.85 and 0.5, respectively. We found that all members of the [WRL] category and about 83\% of the [WRE] objects are red. 
This observed red color can be attributed to either the high interstellar reddening effect along their line-of-sights, the PNe self-reddening due to their high dust concentration content, or the visible light being dominated by its close main sequence binary companion. This astonishing result raises doubts about the commonly held belief that [WR] CS are created as a direct result of the evolved AGBs. 

Toexamine which of these factors is responsible for the apparent reddening of the CSs, we calculated the intrinsic colour (BP-RP)$_0$ of those stars. We obtained the available reddening coefficient E(BP-RP) from the eDR3 database, which only contains reddening coefficients for 21 objects. For the remaining objects, we used Equation \ref{eq:1} to convert E(B- V), which we collected from \citet{Frew16}, to E(BP-RP).
\begin{equation}
E(BP-RP)= 1.289 \times E(B-V) 
\label{eq:1}
\end{equation}
\begingroup
\footnotesize
\begin{longtable}{lllllllllll}
\caption{The color parameters of a sample of our [WR] entire sample. 
Data for the whole sample is available in Table \ref{tab:6a} in the Appendix.}\\
\hline
{\bf serial} & {\bf PN Name} & {\bf G} & {\bf (BP-RP)} &{\bf E(BP-RP)$^1$ } & {\bf (B-V)}  & {\bf E(B-V)} & {\bf E(BP-RP)$^2$}  & {\bf (BP-RP)$_\mathrm{o}$} &{\bf Note} & {\bf Ref.}\\
\hline
\endfirsthead
\multicolumn{11}{c}%
{\tablename\ \thetable\ -- \textit{Continued from previous page}} \\
\hline
{\bf serial} & {\bf PN Name} &{\bf G}& {\bf BP-RP} &{\bf E(BP-RP)$^1$ } &{\bf (B-V)}  &{\bf E(B-V)} &{\bf E(BP-RP)$^2$}  & {\bf (BP-RP)$_\mathrm{o}$} &{\bf Note}&{\bf Ref.}\\
%
\hline
\endhead
\hline
\multicolumn{11}{r}{\textit{Continued on next page}} \\
\endfoot
\hline
\endlastfoot
1  &  PN H 1-62  &  14.4  & 0.77 &  &  & 0.49 & 0.63 & 0.14 &  &  \\
3  &  PN H 1-47  &  15.7  & 1.26 &  & 0.60 & 1.21 & 1.56 & -0.30 &  &  \\
16  &  PN SB 17  &  17.3  & 2.03 & 0.0093 &  &  &  & 2.02 &  &  \\
\hline
\multicolumn{11}{l}{(a) Taken from Gaia eDR3 database, (2) Computed using Eq. \ref{eq:1} (see text).}
\label{tab:6}
\end{longtable}
\endgroup
The results revealed that 61 of the studied 92 [WR]s are intrinsically blue (66\%) while just 34\% (31 items) are red. An additional study was carried out on objects with intrinsic red colour to determine whether the reddening is caused by internal (self-) extinction or close binarity. Only 11 of the objects in Table \ref{tab:6} are close binaries, denoted by the acronym CB. Of this set, we found that WR 72 is the only intrinsically red PN. Unfortunately, we lack a tool to confirm whether the redding of this star is due to its binarity or the self-extinction of its host nebula. For the rest of the sample, the observed reddening might be attributed to the PN internal (self-) extinction.

This argument is supported by previous studies (e.g. \citealt{muth20}) and the analysis of the PNe infrared properties (e.g.,  m$_d$, m$_d$/m$_g$, L$_{IR}$ and infrared excess (IRE)), in \S~\ref{IR}, which showed that the dust content of [WR] nebulae are $\sim$ 3--4 times the non-[WR] nebulae. We found that the dust content in the [WRL] PNe is larger than that in [WRE] group (see Table \ref{tab:5}). Moreover, the dust parameters, m$_d$, m$_d$/m$_g$, and L$_{IR}$, of PNe associated with [WRL] CSs are $\sim$ 4.3, 4.0, and 6.7 times those values for PNe belong to the [WRE] class, respectively.As a result, a plausible explanation for the observed reddening in this subset of [WR]s (30 objects in our sample) is the internal dust extinction of their hosting PNe.

Based on the discussion above, we may infer that, for the studied sample, the key player in the apparent reddening of [WR]s is the ISE (66\%), followed by the internal extinction of the hosting PNe (32\%), while the close binarity has no influence.
\section{Conclusions}
\label{conc}
We analysed a sample of 117 [WR] PNe to investigate the physical and kinematical characteristics of both the [WR] stars and their associated nebulae. In a decreasing order of the T$_{\text{eff}}$ of the [WR] CS, the sample was divided into two groups; early [WRE] and late [WRL]. The location of [WRE] and [WRL] members on the H-R diagram (Fig. \ref{fig:1}) indicates that they have average M$_{\text{final}}$ of 0.61$\pm$0.09 M$_{\odot}$ and 0.58$\pm$0.07 M$_{\odot}$ and a mean t$_{\text{ev}}$ of $11.0\pm3.7 \rm{x}10^3$ yr and $7.9\pm2.7 \rm{x}10^3$ yr, respectively. Generally, [WR] CS are of a slightly higher final mass (0.595$\pm$0.08 M$_{\odot}$) than that of hydrogen poor PG1159 class (0.58$\pm$0.08 M$_{\odot}$), but of a smaller t$_{\text{ev}}$ (9.6$\pm$2.4 \rm{x}10$^3$ yr) than that of their peer PG1159 class (25.5$\pm$5.3 $\times$ 10$^3$ yr; \citealt{Ali21}). The computed mean T$_{\text{eff}}$, L and t$_{\text{dyn}}$ for [WRE] are $110.7\pm5.9$ kK, 7554$\pm$1294 L$_{\odot}$ and $7270\pm1380$ yr while for [WRL] CSs are $35.8\pm2.9$ kK, 5223$\pm$661 L$_{\odot}$ and $5350\pm1170$ yr, respectively.

The mean t$_{\text{dyn}}$ of the entire [WR] sample (= 7270$\pm$1380 yr) is smaller than the mean t$_{\text{ev}}$ ($11012\pm3849$ yr). We may conclude that the computed t$_{\text{dyn}}$, in the current study, are questionable due to the lack of V$_{\text{exp}}$ for a large number of [WR]PNe, in addition to, the assumption of constant velocity during the nebular evolution that we adopted.

While there was no obvious discrimination in nebular shape between [WRE] and [WRL] members, [WRE] PNe had a greater EC than [WRL] objects.
We found that [WRE] PNe are mostly of Peimbert types IIa and IIb whereas [WRL] PNe are dominated by IIb and III types. This result and the mean Z  and V$_{\text{pec}}$ parameters imply that [WRE] PNe belong to the Galactic thin-disk while [WRL] PNe mostly belong to the Galactic thick-disk. In light of the nebular infrared analysis, we obtained higher mean values for the parameters (m$_d$, m$_d$/m$_g$, and L$_{IR}$) in [WRL] than in [WRE] group.

Finally, our data analyses showed that the interstellar extinction is responsible for two-thirds of the red apparent color obtained in the vast majority of [WR] CSs, whereas self-extinction contributed the remaining one-third.


\begin{acknowledgements}
The authors would like to thank the reviewer for the constructive comments that improved the original manuscript.
This work has made use of data from the European Space Agency (ESA) mission
Gaia, processed by the Gaia Data Processing and Analysis Consortium(DPAC). This research has made use
of the SIMBAD database, operated at CDS, Strasbourg, France.
\end{acknowledgements}
\newcommand{\rmxaa}[1]{Revista Mexicana de Astronom{\'\i}a y Astrof{\'\i}sica,}

\newpage
\appendix                  
\section{Full Tables}
\label{app}
\setcounter{table}{0}
\begingroup
\footnotesize
\begin{longtable}{llllcccc}
\caption{A summary of the basic parameters of the studied [WR] sample.}\\
\hline
{\bf serial} & {\bf PN Name} &{\bf [WR] type}&{\bf Gaia eDR3 Source\_ID}& \multicolumn{2}{c}{\bf {Galactic Coordinates}} & \multicolumn{2}{c}{\bf {Equatorial Coordinates}}\\
 & & & & {\bf l($^{\circ}$)} &{\bf b($^{\circ}$) } &{\bf R.A($^{\circ}$)}  &{\bf Dec.($^{\circ}$)}\\
\hline
\endfirsthead
\multicolumn{8}{c}%
{\tablename\ \thetable\ -- \textit{Continued from previous page}} \\
\hline
{\bf serial} & {\bf PN Name} &{\bf [WR] type}&{\bf Gaia eDR3 Source\_ID}& {\bf l($^{\circ}$)} &{\bf b($^{\circ}$) } &{\bf R.A($^{\circ}$)}  &{\bf Dec.($^{\circ}$)}\\
\hline
\endhead
\hline
\multicolumn{8}{r}{\textit{Continued on next page}} \\
\endfoot
\hline
\endlastfoot
1  &  PN H 1$-$ 62  &  [WC 10$-$ 11]  &  4045771305065496832  &  0.0100  &  $-$ 6.8504  &  273.3248  &  $-$ 32.3286  \\ [0.5 ex]
2  &  PN M 2$-$ 20  &  [WC 5$-$ 6]  &  4056503259966429440  &  0.4132  &  $-$ 1.9893  &  268.6055  &  $-$ 29.6023  \\ [0.5 ex]
3  &  PN H 1$-$ 47  &  [WC 11]?  &  4062301564840251520  &  1.2949  &  $-$ 3.0402  &  270.1568  &  $-$ 29.3640  \\ [0.5 ex]
4  &  PN SwSt 1  &  [WC 9]pec  &  4049331244394134912  &  1.5906  &  $-$ 6.7176  &  274.0511  &  $-$ 30.8688  \\ [0.5 ex]
5  &  PN H 1$-$ 55  &  [WC 11]  &  4050131349653595392  &  1.7136  &  $-$ 4.4554  &  271.8106  &  $-$ 29.6902  \\ [0.5 ex]
6  &  PN M 1$-$ 38  &  [WC 5$-$ 6]  &  4050798924999155456  &  2.4831  &  $-$ 3.7454  &  271.5239  &  $-$ 28.6750  \\ [0.5 ex]
7  &  PN K 5$-$ 3  &  [WC 4]  &  4110623760607130624  &  2.6087  &  5.5937  &  262.6717  &  $-$ 23.7501  \\ [0.5 ex]
8  &  Hen 2$-$ 436  &  [WC 4]  &  6744366945682210560  &  4.8712  &  $-$ 22.7265  &  293.0279  &  $-$ 34.2160  \\
\end{longtable}
\endgroup
\footnotesize
\begin{longtable}{llllllll}
\caption{The CS proper motion ({\bf $\mu_\alpha$}, {\bf $\mu_\delta$ }), parallax ({\bf $\varpi$}), and distance (D) as well as the nebular angular radius ({\bf $\theta$}), radial velocity ({\bf V$_r$}) and expansion velocity ({\bf V$_{\text{exp}}$}). Distances marked with the $^{\dag}$ symbol are taken from \citet{Frew16}.}\\
\hline

{\bf Serial} & {\bf $\mu_\alpha \pm$e }& {\bf $\mu_\delta \pm$e} &{\bf $\varpi \pm$e} &{\bf D$\pm$e} &{\bf $\theta$}  &{\bf V$_r \pm$e} &{\bf V$_{\text{exp}}$} \\
 & {\bf (mas/yr)}& {\bf (mas/yr)} &{\bf (mas)} &{\bf pc}&{\bf (arcmin)}  &{\bf (km/s)} &{\bf (km/s)} \\
[0.5 ex] 
\hline
\endfirsthead
\multicolumn{8}{c}%
{\tablename\ \thetable\ -- \textit{Continued from previous page}} \\
\hline
{\bf Serial} & {\bf $\mu_\alpha \pm$e }& {\bf $\mu_\delta \pm$e} &{\bf $\varpi \pm$e} &{\bf D$\pm$e} &{\bf $\theta$}  &{\bf V$_r \pm$e} &{\bf V$_{\text{exp}}$} \\
 & {\bf (mas/yr)}& {\bf (mas/yr)} &{\bf (mas)} &{\bf pc}&{\bf (arcsec)}  &{\bf (km/s)} &{\bf (km/s)} \\[0.5 ex]
\hline
\endhead
\hline
\multicolumn{8}{r}{\textit{Continued on next page}} \\
\endfoot
\hline
\endlastfoot
1 & $-$ 3.681 $\pm$ 0.040 & $-$ 3.597 $\pm$ 0.030 & 0.047 $\pm$ 0.034 &  6970 $\pm$ 2430 & 2.25 & $-$ 86.6 $\pm$ 4.7 & 20 \\  [0.5 ex]
2 & 2.778 $\pm$ 0.268 & $-$ 10.904 $\pm$ 0.172 & $-$ 1.930 $\pm$ 0.240 & $^{\dag}$ 7613 $\pm$ 1670 & 1.88 & 58.7 $\pm$ 15.0 & 20 \\  [0.5 ex]
3 & 1.323 $\pm$ 0.063 & $-$ 7.017 $\pm$ 0.044 & 0.091 $\pm$ 0.050 &  9757 $\pm$ 2139 & 1.25 & 107.5 $\pm$ 3.1 & 20 \\  [0.5 ex]
4 & $-$ 6.238 $\pm$ 0.133 & $-$ 1.720 $\pm$ 0.095 & 0.343 $\pm$ 0.110 &  2914 $\pm$ 934 & 2.70 & $-$ 17.0 $\pm$ 1.6 & 13.75 \\  [0.5 ex]
5 & $-$ 2.445 $\pm$ 0.779 & 1.415 $\pm$ 0.570 & 0.036 $\pm$ 0.860 &  9640 $\pm$ 3340 & 1.45 & $-$ 35.6 $\pm$ 32.6 & 20 \\  [0.5 ex]
6 & $-$ 0.697 $\pm$ 0.031 & $-$ 4.545 $\pm$ 0.023 & 0.120 $\pm$ 0.026 &  8335 $\pm$ 1778 & 1.75 & $-$ 82.8 $\pm$ 9.1 & 20 \\  [0.5 ex]
7 & $-$ 4.363 $\pm$ 0.539 & $-$ 9.627 $\pm$ 0.365 & 0.200 $\pm$ 0.337 &  4540 $\pm$ 1000 & 6.50 & 144.9 $\pm$ 7.1 & 20 \\  [0.5 ex]
8 & $-$ 0.699 $\pm$ 0.454 & $-$ 1.563 $\pm$ 0.380 & $-$ 0.521 $\pm$ 0.460 &     & 0.30 &    & 14 \\
\end{longtable}
\begin{landscape}
\begingroup
\footnotesize
\begin{longtable}{lllllllllllll}
\caption{The effective temperature (T$_{\text{eff}}$) and luminosity (L) of [WR] CSs in logarithmic scale, final CS mass ({\bf M$_{\text{final}}$}) and evolutionary age ({\bf t$_{\text{ev}}$}). Further, the table includes the space velocity components (U, V, W) and radial peculiar velocity ({\bf V$_{\text{pec}}$}) of the [WR] stars.  The sample is arranged into two groups, Early: [WRE] and late: [WRL]. The table is limited to objects with available data.}\\
\hline
{\bf serial} & {\bf PN Name} & {\bf [WR] type}& \multicolumn{4}{c}{\bf Physical properties}&&\multicolumn{5}{c}{\bf Kinematical Parameters}\\

&&&{\bf log T$_{\text{eff}}$} & {\bf log L}& {\bf M$_{\text{final}}$ }& {\bf t$_{\text{ev}}$} & &{\bf U$_{\text{LSR}} \pm$ e} &{\bf V$_{\text{LSR}} \pm$ e}& {\bf W$_{\text{LSR}} \pm$ e}& {\bf V$_s \pm$ e} &{\bf V$_{\text{pec}} \pm$ e}\\
 & &&{\bf (K)}& {\bf (L$_{\odot}$)} &{\bf (M$_{\odot}$)} &{\bf (yr)}&& \multicolumn{5}{c}{\bf km/s}\\ [0.5 ex] 
\hline
\endfirsthead
\multicolumn{12}{c}%
{\tablename\ \thetable\ -- \textit{Continued from previous page}} \\
\hline
{\bf serial} & {\bf PN Name} & {\bf [WR] type}&{\bf log T$_{\text{eff}}$} & {\bf log L }& {\bf M$_{\text{final}}$ }& {\bf t$_{\text{ev}}$} & &{\bf U$_{\text{LSR}} \pm$ e} &{\bf V$_{\text{LSR}} \pm$ e}& {\bf W$_{\text{LSR}} \pm$ e}& {\bf V$_s \pm$ e} &{\bf V$_{\text{pec}} \pm$ e} \\
%
\hline
\endhead
\hline
\multicolumn{12}{r}{\textit{Continued on next page}} \\
\endfoot
\hline
\endlastfoot
\multicolumn{12}{c}{\bf [WRE] }  \\ [0.5 ex] \hline

 7 & PN K 5$-$ 3 & [WC 4] &  &  &  &  & & 168.7 $\pm$ 8.3 & 8.4 $\pm$ 2.0 & $-$ 13.8 $\pm$ 8.5 & 271 $\pm$ 5.4 & 127 $\pm$ 7.6 \\  [0.5 ex]
 11 & PN M 2$-$ 31 & [WC 4] &  &  &  &  & &    &    &    &    & 89 $\pm$ 14.1 \\  [0.5 ex]
 12 & PN M 3$-$ 15 & [WC 4] &  &  &  &  & & 117.5 $\pm$ 2.7 & 185.4 $\pm$ 52.5 & $-$ 61.7 $\pm$ 14.4 & 137 $\pm$ 7.3 & 47 $\pm$ 11.5 \\  [0.5 ex]
 21 & PN M  1$-$ 60 & [WC 4] & 4.87 & 3.93 & 0.59 & 772 & &    &    &    &    & 79 $\pm$ 38.2 \\  [0.5 ex]
 27 & PN M 4$-$ 13 & [WC 4] &  &  &  &  & & 149.6 $\pm$ 32.8 & 103.9 $\pm$ 23.2 & $-$ 1.1 $\pm$ 0.4 & 189 $\pm$ 30.4 & 31 $\pm$ 12.4 \\  [0.5 ex]
 48 & PN Pe 1$-$ 1 & [WC 5] &  &  &  &  & &    &    &    &    & 11 $\pm$ 3.1 \\  [0.5 ex]
 51 & WRAY 16$-$ 92 & [WC 4] &  &  &  &  & &    &    &    &    & 69 $\pm$ 4.4 \\
\end{longtable}
\endgroup
\end{landscape}
\begingroup
\footnotesize
\begin{longtable}{lllllllll}
\caption{The physical nebular parameters surrounding the [WR]s. The PN scale height (Z), radius (R), 5GHz flux ({\bf F$_{\text{5GHz}}$}), ionized mass ({\bf m$_{ion}$}), optical depth ({\bf $\tau$}), and radio surface temperature ({\bf T$_B$}).}\\
\hline
{\bf serial} & {\bf PN Name} &{\bf Z $\pm$ e }& {\bf R $\pm$ e} &{\bf t$_{\text{dyn}} \pm$ e}&{\bf F$_{\text{5GHz}}\pm$ e} & {\bf m$_{ion}\pm$ e} & {\bf $\tau$} &{\bf T$_B$} \\
             &               &{\bf (pc)}      & {\bf (pc)}      &{\bf (yr/10$^3$)}& {\bf (mJy)}& (M$_{\odot}$) & &{\bf (K)}\\ [0.5 ex]
\hline
\endfirsthead
\multicolumn{9}{c}%
{\tablename\ \thetable\ -- \textit{Continued from previous page}} \\
\hline
{\bf serial} & {\bf PN Name} &{\bf Z $\pm$ e }& {\bf R $\pm$ e} &{\bf t$_{\text{dyn}} \pm$ e}&{\bf 5GHz $\pm$ e}& {\bf m$_{ion} \pm$ e} & {\bf $\tau$} &{\bf T$_B$} \\
             &               &{\bf (pc)}      & {\bf (pc)}      &{\bf (yr/10$^3$)}& {\bf (mJy)}& & &{\bf (K)} \\ [0.5 ex]
\hline
\endhead
\hline
\multicolumn{7}{r}{\textit{Continued on next page}} \\
\endfoot
\endlastfoot
\multicolumn{9}{c}{\bf [WRE] }  \\ [0.5 ex] \hline
 7 & PN K 5$-$ 3 & 443 $\pm$ 97 & 0.143 $\pm$ 0.032 & 7.00E+03 $\pm$ 1.54E+03 &    &    &  &  \\ [0.5 ex]
 11 & PN M 2$-$ 31 & 383 $\pm$ 75 & 0.057 $\pm$ 0.011 & 2.77E+03 $\pm$ 5.44E+02 & 51.0 $\pm$ 10.20 & 0.069 $\pm$ 0.036 & 2.46 & 252 \\ [0.5 ex]
 12 & PN M 3$-$ 15 & 391 $\pm$ 81 & 0.057 $\pm$ 0.012 & 2.78E+03 $\pm$ 5.75E+02 &    &    &  &  \\ [0.5 ex]
 21 & PN M  1$-$ 60 & 568 $\pm$ 117 & 0.044 $\pm$ 0.009 & 2.13E+03 $\pm$ 4.40E+02 & 50.0 $\pm$ 11.00 & 0.055 $\pm$ 0.030 & 2.10 & 586 \\ [0.5 ex]
 22 & PMR 7 & 699 $\pm$ 14 & 0.348 $\pm$ 0.007 & 1.70E+04 $\pm$ 3.44E+02 &    &    &    \\ [0.5 ex]
 27 & PN M 4$-$ 13 & 159 $\pm$ 34 & 0.062 $\pm$ 0.013 & 3.04E+03 $\pm$ 6.40E+02 & 69.0 $\pm$ 13.80 & 0.070 $\pm$ 0.038 & 2.66 & 161 \\ [0.5 ex]
 48 & PN Pe 1$-$ 1 & 135 $\pm$ 29 & 0.037 $\pm$ 0.008 & 1.83E+03 $\pm$ 3.99E+02 &    &    &    \\
\end{longtable}
\endgroup

\begingroup
\footnotesize
\begin{longtable}{llllllllll}
\caption{The excitation class (EC), Peimbert (Peim.) and morphological (mor.) types, dust temperature ({\bf T$_{\text{dust}}$} ), dust mass ({\bf m$_{\text{dust}}$}), dust/gas mass ratio ({\bf m$_{\text{dust}}$/m$_{\text{gas}}$ }), and infrared luminosity ({\bf L$_{\text{IF}}$}) of [WR]PNe.}\\
\hline
{\bf serial} & {\bf PN Name} & {\bf EC } & {\bf Peim.} & {\bf Mor.}  && {\bf T$_{\text{dust}}$} & {\bf m$_{\text{dust}} \pm$ e}& {\bf m$_{\text{dust}}$/m$_{\text{gas}} \pm$ e}& {\bf L$_{\text{IF}} \pm$ e} \\
& &  & {\bf Type} & {\bf Type}  & &(K) & ($10^{-4}$\,M$_{\odot}$)&($10^{-3}$) &  (L$_{\odot}$) \\ [0.5ex]
\hline
\endfirsthead
\multicolumn{10}{c}%
{\tablename\ \thetable\ -- \textit{Continued from previous page}} \\
\hline
{\bf serial} & {\bf PN Name} & {\bf EC } & {\bf Peim.} & {\bf Mor. }  && {\bf T$_{\text{dust}}$} & {\bf m$_{\text{dust}} \pm$ e}& {\bf m$_{\text{dust}}$/m$_{\text{gas}} \pm$ e}& {\bf L$_{\text{IF}} \pm$ e} \\
& &  & {\bf Type} & {\bf type} && (K) & (M$_{\odot}$)& &  (L$_{\odot}$) \\ [0.5ex]
\hline
\endhead
\hline
\multicolumn{10}{r}{\textit{Continued on next page}} \\
\endfoot
\endlastfoot
\multicolumn{10}{c}{\bf [WRE] }  \\ [0.5 ex] \hline
 7 & PN K 5$-$ 3 & 11.0 & IIb & R & &  &    &    &    \\  [0.5 ex]
 8 & Hen 2$-$ 436 & 3.0 &  & S & & 115 & 1.84 $\pm$ 0.75 &    & 2934 $\pm$ 1250 \\  [0.5 ex]
 11 & PN M 2$-$ 31 & 2.0 &  & E & & 86 & 1.73 $\pm$ 0.72 & 2.9 $\pm$ 0.5 & 642 $\pm$ 340 \\  [0.5 ex]
 12 & PN M 3$-$ 15 & 6.0 & IIa & B & & 89 & 2.01 $\pm$ 0.89 & 2.6 $\pm$ 0.5 & 888 $\pm$ 385 \\  [0.5 ex]
 20 & IRAS 18249$-$ 1506 &  &  & R & & 77 &    &    &    \\  [0.5 ex]
 21 & PN M  1$-$ 60 & 4.0 & IIa & B & & 87 & 3.41 $\pm$ 0.55 & 4.9 $\pm$ 0.6 & 1348 $\pm$ 741 \\  [0.5 ex]
 22 & PMR 7 &  &  & E & &  &    &    &    \\
\end{longtable}
\endgroup
\begingroup
\footnotesize
\begin{longtable}{lllllllllll}

\caption{Summary of the colour parameters of our [WR] CSs sample. In the notes column, the abbreviation CB refers to close binary.}\\
\hline
{\bf serial} & {\bf PN Name} & {\bf G} & {\bf (BP-RP)} &{\bf E(BP-RP)$^a$ } & {\bf (B-V)}  & {\bf E(B-V)} & {\bf E(BP-RP)$^b$}  & {\bf (BP-RP)$_\mathrm{o}$} &{\bf Note} & {\bf Ref.}\\
\hline
\endfirsthead
\multicolumn{11}{c}%
{\tablename\ \thetable\ -- \textit{Continued from previous page}} \\
\hline
{\bf serial} & {\bf PN Name} &{\bf G}& {\bf BP-RP} &{\bf E(BP-RP)$^a$ } &{\bf (B-V)}  &{\bf E(B-V)} &{\bf E(BP-RP)$^b$}  & {\bf (BP-RP)$_\mathrm{o}$} &{\bf Note}&{\bf Ref.}\\
%
\hline
\endhead
\hline
\multicolumn{11}{r}{\textit{Continued on next page}} \\
\endfoot
\hline
\endlastfoot

1  &  PN H 1$-$ 62  &  14.4  & 0.77 &  &  & 0.49 & 0.63 & 0.14 &  &  \\
2  &  PN M 2$-$ 20  &  15.9  & 1.66 &  &  & 1.29 & 1.66 & 0.00 &  &  \\
3  &  PN H 1$-$ 47  &  15.7  & 1.26 &  & 0.60 & 1.21 & 1.56 & $-$ 0.30 &  &  \\
4  &  PN SwSt 1  &  11.8  & 0.85 &  & 0.01 & 0.24 & 0.31 & 0.54 &  &  \\
5  &  PN H 1$-$ 55  &  16.6  & 1.13 &  & 0.46 & 0.79 & 1.02 & 0.11 &  &  \\
6  &  PN M 1$-$ 38  &  14.1  & 0.93 &  & 0.45 & 0.63 & 0.81 & 0.12 &  &  \\
7  &  PN K 5$-$ 3  &  19.2  & 1.1 &  &  & 1.07 & 1.38 & $-$ 0.28 &  &  \\
8  &  Hen 2$-$ 436  &  14.9  & $-$ 0.06 &  &  & 0.11 & 0.14 & $-$ 0.20 &  &  \\
\end{longtable}
\endgroup

\end{document}